\documentclass[dvipdfmx]{jps-cp}
\usepackage{txfonts} %Please comment out this line unless the txfonts package is availabe in your LaTeX system.

\usepackage{amsmath}
\usepackage{bm}
\usepackage{physics}
\usepackage[dvipdfmx]{graphicx}
\usepackage[dvipdfmx]{color}
\usepackage{here}
\usepackage{xcolor}
\usepackage{listings,jvlisting}

\title{Possible condensation of Cooper triples}

\author{Sora \textsc{Akagami}$^{1}$, Hiroyuki \textsc{Tajima}$^{2}$, and Kei \textsc{Iida}$^{1}$}

\inst{$^{1}$Department of Mathematics and Physics, Kochi University, Kochi 780-8520, Japan \\
$^{2}$Department of Physics, The University of Tokyo, Tokyo 113-0033, Japan}

\email{b193p001@s.kochi-u.ac.jp}

\recdate{July 7, 2022}

\abst{
We theoretically discuss the possible condensation of Cooper triples, which correspond to a three-body version of Cooper pairs, in three-component Fermi systems with three-body attractive interactions.
A macroscopic number of Cooper triples can occupy a zero center-of-mass momentum state in the presence of a Fermi surface of constituent particles, even though the three-body operator exhibits anti-commutation relation associated with the Fermi--Dirac statistics.
Such a condensation with internal degrees of freedom is similar to bosonization in a system of {\it infinite}-component fermions.
We propose a variational wave function for condensed Cooper triples and show that in the ground state, the condensed state is energetically favored compared to the normal state. 
Also, we discuss effects of the Fermi-surface distortion in a lattice system described by a three-component Hubbard model.
}

\kword{Cooper triple, fermionic condensates, Variational approach}

\begin{document}
\setlength\textfloatsep{0pt}
\maketitle
\section{Introduction}
The Bardeen Cooper-Schrieffer (BCS) theory~\cite{BCS},
which explains the microscopic mechanism of Cooper-pair formation and its condensation due to attractive two-body interactions between two different species,
has had a major impact not only in condensed matter physics but also nuclear and particle physics.
The Cooper-pair condensation has been understood as a bosonic-like condensation in which many Cooper pairs occupy a zero center-of-mass (COM) momentum state.

In recent years, the concept of Cooper pairs has been generalized to cluster states consisting of three or more particles, such as Cooper triples \cite{Hammer2012, Tajima2021} and Cooper quartets \cite{Kamei,Baran,Guo2022}.
While these kinds of states have been widely discussed in nuclear and condensed matter systems,
ultracold atomic gases may provide a promising platform to investigate the existence of nontrivial Cooper multiplets.
In fact, the momentum distribution of fermions forming Cooper pairs has been directly observed in a recent state-of-the-art experiment~\cite{Holten2022}.
Moreover, multi-component Fermi gases can be realized by using different hyperfine states of ultracold atoms (e.g., $^6$Li~\cite{Ohara2011}, $^{173}$Yb~\cite{Schafer2020}).
The long-range order in a multi-component attractive Hubbard model has also been discussed theoretically~\cite{Yoshida2021,Nakagawa2022}. 
The realization of a three-body interaction, which is deeply related to the formation of Cooper triples, has also been proposed
in cold atomic systems~\cite{Petrov1,Valiente,Akagami2021,Tajima2021}.
The three-body condensation has also been predicted in non-equilibrium bosonic systems~\cite{Musolino2022}.
%In this paper, we would like to focus on Cooper triples.
%In three-component Fermi systems, the formation of Cooper triples, corresponding to three-body versions of Cooper pairs, was studied in the presence of two- and three-body interactions.

\begin{figure}[t]
    \begin{center}
    \includegraphics[width=9cm]{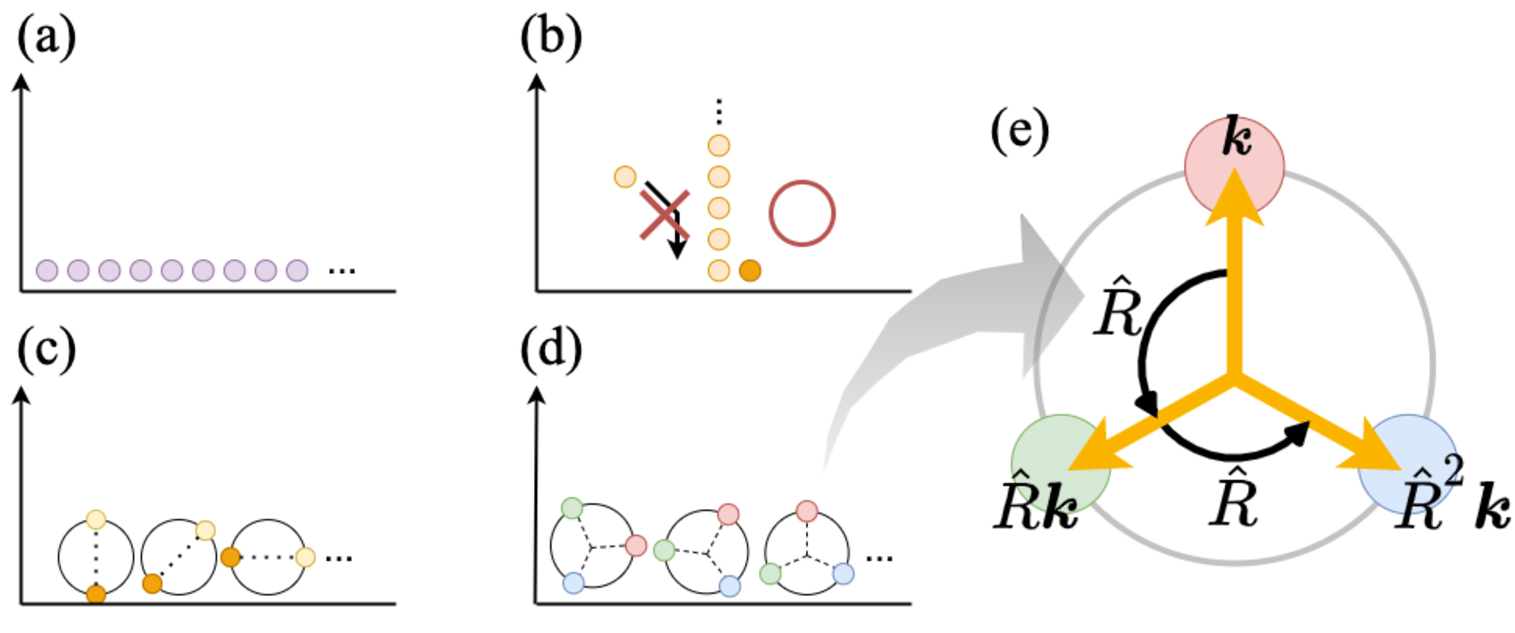}
    \caption{
    Schematic diagrams of the occupation of particles and bound states at zero temperature for the Bose--Einstein condensation (a), the Fermi degeneracy (b), the Cooper pairs' condensation (c), and the Cooper triples' condensation (d). Different colors represent different components.
    %, with only purple  representing boson.
    (e) The in-plane configuration of the three fermions of momentum ${\bm k}, \hat{R}{\bm k}$, and $\hat{R}^2{\bm k}$, which we call three-point-type configuration (TPTC). The $\hat{R}$ represents the $2\pi/3$ radial rotation operator in a given momentum plane and hence ensures zero COM momentum; ${\bm k}+\hat{R}{\bm k}+\hat{R}^2{\bm k}={\bm 0}$.
    %(this configuration is the TPTC). 
    % For \cite{Akagami2021}
    % [S. Akagami, H.Tajima, and K.Iida, Phys. Rev. A {\bf 104}, L041302 (2021). Copyright (2021) by the American Physical Society. https://doi.org/10.1103/PhysRevA.104.L041302]
    }
    \label{fig:configuration}
    \end{center}
\end{figure}

In this paper,
as a consequence of the formation of a macroscopic number of Cooper triples,
we show that Cooper triples may condense
at zero COM momentum
in a three-component Fermi system 
by utilizing internal degrees of freedom such as relative momenta among constituent particles (see Fig.~\ref{fig:configuration}).
For convenience,
we schematically compare this nontrivial condensation with the conventional quantum degenerate states of
point-like fermions/bosons and Cooper pairs.
As in the case of Cooper pairs with nonzero relative momentum between two fermions (in contrast to tightly bound dimers), Cooper triples are regarded as loosely bound three-body states where constituent fermions possess nonzero momenta being close to the Fermi momentum.
In the following, we call the configuration of three fermions near the Fermi surface (FS) with zero COM momentum as the three-point-type configuration~(TPTC). 
A numerous number of TPTC 
(i.e., condensation at zero COM momentum) 
can be realized by the infinitesimal rotation of the momenta of constituent fermions,
without conflicting with the Pauli-exclusion principle of three-body states.
The condensation with such additional internal degrees of freedom can be regarded as an analogy of bosonization in SU($N$) fermions at large $N$ limit~\cite{Song2020}.
We derive the ground-state energy within the variational theory and show that the Cooper triple state is energetically more favorable than the normal state.
While a gas system exhibits an isotropic FS and one can consider the rotational symmetry in the momentum space, the FS in lattice systems becomes anisotropic due to the lattice geometry. 
In this paper, stepping further from our previous work~\cite{Akagami2021},
we examine effects of FS anisotropy on the realization of TPTC in a two-dimensional square-lattice Hubbard model with three-body attraction.
Thorough this paper, we take $\hbar=k_{\rm B}=1$ and the system volume is take to be unity.

\section{Cooper triple condensation}
%We consider non-relativistic three-component $({\rm r,g,b})$ fermions of equal mass $m$ with attractive three-body interaction. 
%The various system with three-body interactions has been discussed in ultracold atomic gases.
%Previous studies \cite{Hammer2012} have predicted that Cooper triples are more dominant than Cooper pairs above a certain threshold strength, even when there is only an attractive two-body interaction between the three-components.
%In a paper we previously submitted\cite{Akagami2021}, we showed that Cooper triples can form in systems where three-body interactions exist, by considering a three-particle version of the Cooper problem.
%We have also proposed a way to achieve an attractive three-body interaction in multi-component ultracold atomic gases (e.g. ${}^6{\rm Li}$, ${}^{173}{\rm Yb}$) using a background medium gas.
%In this paper, for simplicity, we assume that the two-body interaction vanishes and that only a weakly attractive three-body interaction exists. We also consider the system in which the chemical potentials are the same, i.e. $\mu_{\rm r}=\mu_{\rm g}=\mu_{\rm b}\equiv\mu$.
We consider three-component fermions with three-body attractive interaction loaded in the two-dimensional square lattice, where we assume that the three-body scattering occurs only in the vicinity of the FS at sufficiently low temperatures and weak interactions, as described by the Hamiltonian
\begin{align}
    \hat{H}
    =\sum_{\gamma={\rm r,g,b}}\sum_{\bm k}\xi_{\bm k}\hat{n}_{{\bm k},\gamma}
    +\sum_{{\bm k},\hat{R}}\sum_{{\bm k}',\hat{R}'}U_{{\bm k},{\bm k}'}\hat{F}^\dag\hspace{-5pt}{}_{{\bm k}',\hat{R}'}\hat{F}{}_{{\bm k},\hat{R}}.
    \label{eq:bulk_hamiltonian}
\end{align}
Here, $\xi_{\bm k}$ is the kinetic energy of a fermion with the lattice momentum $\bm{k}=(k_x,k_y)$.
While the kinetic energy in a gas system is given by the parabolic dispersion $\xi_{\bm k}^{\rm gas}=k^2/(2m)-\mu$ (where $\mu$ is the chemical potential),
it is given by $\xi_{\bm k}=-2t\cos k_x-2t\cos k_y-\mu$ in the two-dimensional square lattice where $t$ is the hopping parameter and
the lattice constant is taken to be unity.
However, the low-momentum expansion of $\xi_{\bm k}$ reproduces the parabolic dispersion as $\xi_{\bm k}=tk^2-\mu+O(k^4)$ where the constant energy shift is omitted.
First, for simplicity, we consider the low-filling case where the FS can be assumed to be isotropic in the following (effects of FS anisotropy will be discussed in the next section).
$\hat{n}_{{\bm k},\gamma}=\hat{c}^\dag\hspace{-5pt}{}_{{\bm k},\gamma}\hat{c}_{{\bm k},\gamma}$ is the single-particle number operator with the annihilation (creation) operator $\hat{c}_{{\bm k},\gamma}^{(\dag)}$ for fermions of hyperfine state $\gamma={\rm r}, {\rm g}, {\rm b}$. 
In this paper, we consider the equally-populated three-component mixture in the thermodynamic limit.

The second term in Eq.\ \eqref{eq:bulk_hamiltonian} denotes the three-body interaction with a contact-type coupling constant $U_{{\bm k},{\bm k}'}\ (<0)$. $\hat{F}_{{\bm k},\hat{R}}^{(\dag)}$ is the three-body annihilation (creation) operator defined in terms of $\hat{c}_{{\bm k},\gamma}$ as
\begin{align}
    \hat{F}^\dag\hspace{-5pt}{}_{{\bm k},\hat{R}}=\hat{c}^\dag\hspace{-5pt}{}_{{\bm k},{\rm r}}\hat{c}^\dag\hspace{-5pt}{}_{\hat{R}{\bm k},{\rm g}}\hat{c}^\dag\hspace{-5pt}{}_{\hat{R}^2{\bm k},{\rm b}},\quad
    \hat{F}{}_{{\bm k},\hat{R}}=\hat{c}{}_{\hat{R}^2{\bm k},{\rm b}}\hat{c}{}_{\hat{R}{\bm k},{\rm g}}\hat{c}{}_{{\bm k},{\rm r}}.
\end{align}
$\hat{R}$ is the $2\pi/3$-rotation operator acting on the lattice momentum in the COM frame.
For each of the three fermions near the FS, the absolute value of the momentum $|{\bm k}|$, $|{\hat{R}{\bm k}}|$, and $|{\hat{R}^2{\bm k}}|$ are close to the Fermi momentum $k_{\rm F}$, while the COM momentum remains zero 
(i.e., $\bm{k}+\hat{R}\bm{k}+\hat{R}^2\bm{k}=\bm{0}$).
${\bm k}$, $\hat{R}{\bm k}$, and $\hat{R}^2{\bm k}$ are on the same plane with $2\pi/3$ and $4\pi/3$ rotations in the momentum space as shown in Fig.~\ref{fig:configuration}.
Therefore, by fixing ${\bm k}$ and the rotational axis, the remaining two momenta, $\hat{R}{\bm k}$ and $\hat{R}^2{\bm k}$, are automatically determined.
$\hat{F}_{{\bm k},\hat{R}}^{(\dag)}$ satisfies the following anti-commutation relations
\begin{align}
    \Bigl\{\hat{F}_{{\bm k},\hat{R}},\hat{F}_{{\bm k}',\hat{R}'}\Bigr\}
    &=\Bigl\{\hat{F}^\dag\hspace{-5pt}{}_{{\bm k},\hat{R}},\hat{F}^\dag\hspace{-5pt}{}_{{\bm k}',\hat{R}'}\Bigr\}=\hat{0},\label{eq:acomm1}\\
    \Bigl\{\hat{F}_{{\bm k},\hat{R}},\hat{F}^\dag\hspace{-5pt}{}_{{\bm k}',\hat{R}'}\Bigr\}
    &=\delta_{{\bm k},{\bm k}'}\delta_{\hat{R}{\bm k},\hat{R}'{\bm k}'}\delta_{\hat{R}^2{\bm k},\hat{R}'^2{\bm k}'}
    \Bigl[\Bigl(\hat{1}-\hat{n}_{{\bm k},{\rm r}}\Bigr)\Bigl(\hat{1}-\hat{n}_{\hat{R}{\bm k},{\rm g}}\Bigr)\Bigl(\hat{1}-\hat{n}_{\hat{R}^2{\bm k},{\rm b}}\Bigr)+\hat{n}_{{\bm k},{\rm r}}\hat{n}_{\hat{R}{\bm k},{\rm g}}\hat{n}_{\hat{R}^2{\bm k},{\rm b}}\Bigr].\label{eq:acomm2}
\end{align}
%Next, we consider different three-particle fermion configurations as shown in Fig.\ \ref{fig:configuration} and discuss how the condensation of Cooper triples can occur.
%%%%%Configurationsの図
% \begin{figure}[H]
%     \begin{center}
%     \includegraphics[width=7cm]{Fig2.eps}
%     \caption{These are schematic diagrams of the occupation of particles and bound states at zero temperature. Different colors represent different components. Only purple represents boson, all other colors represent fermion. Dotted lines represent two- or three-body interactions.
%     The top left figure represents the Bose-Einstein condensation. The top right figure represents Fermi degeneracy. The bottom left figure represents Cooper pairs condensate; two fermions of different species occupy a single state with zero center-of-mass momentum. The bottom right figure represents Cooper triples condensation; three fermions of different species occupy a single state with zero center-of-mass momentum.
%     }
%     \label{fig:condensations}
%     \end{center}
% \end{figure}
In particular, from the anti-commutation relation associated with $\hat{c}_{{\bm k},\gamma}^\dag$, one finds
\begin{align}
\label{eq:5}
    \hat{F}^\dag\hspace{-5pt}{}_{{\bm k},\hat{R}}\hat{F}^\dag\hspace{-5pt}{}_{{\bm k}',\hat{R}'}\ket{0}\neq0\quad
    ({\bm k}\neq{\bm k}',\ \hat{R}{\bm k}\neq\hat{R}'{\bm k}',\ \hat{R}^2{\bm k}\neq\hat{R}'^2{\bm k}'),
    %6体状態と以前勘違いされたことがありますが、ここで3つだとk\neq k' \neq k''と書いてもk=k''の可能性を排除できず面倒なので2つにしておいた方がよさそう
    %\hat{F}_{{\bm k},\hat{R}}^\dag\hat{F}_{{\bm k}',\hat{R}'}^\dag\hat{F}_{{\bm k}'',\hat{R}''}^\dag\ket{0}\neq0\quad
    %({\bm k}\neq{\bm k}'\neq{\bm k}'',\ \hat{R}{\bm k}\neq\hat{R}'{\bm k}'\neq\hat{R}''{\bm k}'',\ \hat{R}^2{\bm k}\neq\hat{R}'^2{\bm k}'\neq\hat{R}''^2{\bm k}''),
\end{align}
where $\ket{0}$ is the normalized vacuum state.
Eq.~(\ref{eq:5}) indicates that two three-body states with zero COM momentum can coexist as long as the momenta of a certain component are different. %is not the same as the other ones.
Since there are an infinite number of ways to take ${\bm k}$, one can find a macroscopic number of TPTC in the presence of FS.
%This appears as if bosonization of the {\it infinite}-component fermions \cite{Song2020} had occurred.
Thus, even though the Cooper triples obey the Fermi--Dirac statistics, these three-body states can macroscopically occupy a zero COM momentum state and form the condensate.
%Relative momentum is a key to this condensation, which can be thought of in the same way as the condensation of Cooper pairs, as shown in Fig. \ref{fig:configuration}.
We recall that the relative momenta of constituent fermions plays a key role in the condensation of Cooper triples as shown in Fig.~\ref{fig:configuration}.
Such internal degrees of freedom also exist in usual Cooper pairs and makes a difference between Cooper pairs and tightly bound dimers in terms of the spatial structure of wave functions~\cite{Ohashi2020}.
%Although the Cooper pair condensation is a bosonic-like condensation, it is not strictly a Bose-Einstein condensation.
%However, by viewing condensation in relation to relative momentum, it becomes easier to view Cooper pair condensation and Cooper triple condensation as the same mechanism. 

Using the variational wave function inspired by the BCS theory~\cite{Akagami2021},
we show that the condensed state of Cooper triples are favored over the normal phase.
We consider the variational wave function
\begin{align}
    \ket{\Psi_T}=\prod_{\bm k}(u_{\bm k}+v_{\bm k}\hat{F}^\dag\hspace{-5pt}{}_{{\bm k},\hat{R}_0})\ket{0},
    \label{eq:wave_function}
\end{align}
where $u_{\bm k}$ and $v_{\bm k}$ are variational parameters satisfying the normalization condition $\abs{u_{\bm k}}^2+\abs{v_{\bm k}}^2=1$ 
%it is
(associated with $\braket{\Psi_T}=1$). 
% \blue{
% This variational wave function is a mixture of fermionic states (odd numbers of $\hat{F}^\dag\hspace{-5pt}{}_{{\bm k},\hat{R}}$) or bosonic  states (even numbers of $\hat{F}^\dag\hspace{-5pt}{}_{{\bm k},\hat{R}}$) with different numbers of particles. 
% In any case, it is a variational wave function that describes the state in which countless $\hat{F}^\dag\hspace{-5pt}{}_{{\bm k},\hat{R}}$ are computed, i.e., the Cooper triples are condensed.
% }
We note that the state \eqref{eq:wave_function} breaks the $U$(1) gauge symmetry via the superposition of the Cooper triples.
Here, we set $\hat{R}_0$ such that if the momentum ${\bm k}$ of a fermion with $\gamma={\rm r}$ is in the direction of 
% ${\bm e}_1=\mqty(0\\0\\1)$, 
${\bm e}_1=(0, 0)^{\rm T}$,
then the momentum $\hat{R}_0{\bm k}$ 
% (corresponding to a fermion with $\gamma={\rm g}$) 
($\gamma={\rm g}$) 
parallels
%is in 
% ${\bm e}_2=\mqty(\sqrt{3}/2\\0\\-1/2)$
${\bm e}_2=(-\sqrt{3}/2, -1/2)^{\rm T}$
and the remaining one $\hat{R}_0^2{\bm k}$ ($\gamma={\rm b}$) 
parallels
%is in
% ${\bm e}_3=\mqty(-\sqrt{3}/2\\0\\-1/2)$.
${\bm e}_3=(\sqrt{3}/2, -1/2)^{\rm T}$.
%We note that 
{For other ${\bm k}$'s},
${\bm k}$, $\hat{R}_0{\bm k}$, and $\hat{R}_0^2{\bm k}$ are oriented in the direction of $V{\bm e}_1$, $V{\bm e}_2$, and $V{\bm e}_3$, respectively, 
with an appropriate rotation matrix $V$.
In this sense,
although this choice of $\hat{R}_0$ leads to a specific orientation in momentum space, 
%however, 
Eq.~\eqref{eq:wave_function} can have any momenta of each component picked up once in the infinite product of $\bm{k}$. 
This fact is also important for reproducing the normal Fermi degenerate state $|{\rm FS}\rangle=\prod_{|\bm{k}|\leq k_{\rm F},\gamma}\hat{c}^\dag\hspace{-5pt}{}_{\bm{k},\gamma}|0\rangle$ within the variational space of Eq.~\eqref{eq:wave_function}.
%%%here
% Note that one can consider another $\hat{R}_0$ 
% % by setting 
% % % ${\bm e}_2=\mqty(\sqrt{3}\cos{\theta}/2\\\sqrt{3}\sin{\theta}/2\\-1/2)$
% % ${\bm e}_2=(\sqrt{3}\cos{\theta}/2,\sqrt{3}\sin{\theta}/2,-1/2)^{\rm T}$ and 
% % % ${\bm e}_3=\mqty(-\sqrt{3}\cos{\theta}/2\\-\sqrt{3}\sin{\theta}/2\\-1/2)$ 
% % ${\bm e}_3=(-\sqrt{3}\cos{\theta}/2,-\sqrt{3}\sin{\theta}/2,-1/2)^{\rm T}$  
% % with $\theta\neq0$, i.e., 
% by rotating the momentum plane on which a Cooper triple with ${\bm k}$ in the direction of ${\bm e}_1$ resides by $\theta$ for ${\bm e}_1$. 
Note that one can consider a different $\hat{R}_0$ by considering different momentum plane.
% The resultant state is degenerate from the original state.
The resultant state is degenerate with the original state.
Such degeneracy is similar to the case of different gauge orientations, and the variational parameters can be regarded as independent of the choice of $\hat{R}_0$ as well as the gauge.
In the BCS state, only ${\bm k}$ determines the relative momentum of a Cooper pair, but in the present state, not only ${\bm k}$ but also $\hat{R}_0$ characterizes the relative momentum between two of three fermions forming a Cooper triple.

%As a more sophisticated variational wave function, we can also consider the superposition of the state given by Eq. \eqref{eq:wave_function} with the states at various $\hat{R}_0$.
%In such a superposition, the variational parameter space is larger and the energy of the ground-state may be lower than in \eqref{eq:wave_function}.
%For our purposes, however, it is sufficient to show that the formation of the Cooper triple considered in Eq. \eqref{eq:wave_function} results in a lower energy of the ground-state of the Cooper triple than the normal state.
%Since we are interested in the case with infinitely weak attractive three-body interaction in Eq. \eqref{eq:bulk_hamiltonian}, the zero center-of-mass momentum configuration of Cooper triple is restricted to the TPTC on the Fermi surface.
%In the strong-coupling case, it is possible to construct a variational wave function consisting of the creation operator $\hat{F}^\dag$ of a configuration of three fermions of different species of zero center-of-mass momentum, but it has been difficult to find a variational wave function that can describe a common Fermi surface for all three-components, not just the condensed state.
%This is because, for $\hat{F}^\dag$, if we choose a particular triangle with a central mass at the origin and different lengths from the origin in momentum space, we can not describe the Fermi sphere and the transition to the condensed state.

Using Eq.~\eqref{eq:wave_function},
we evaluate the ground-state energy $E_0=\bra{\Psi_T}\hat{H}\ket{\Psi_T}$ as
\begin{align}
    E_0=\sum_{\bm k}3\xi_{\bm k}\abs{v_{\bm k}}^2
    +\sum_{\bm k}\sum_{{\bm k}'}U_{{\bm k},{\bm k}'}v_{\bm k}^\ast v_{{\bm k}'}u_{{\bm k}'}^\ast u_{\bm k}.
    \label{eq:Energy_in_Variational_method}
\end{align}
The minimization of $E_0$ with respect to the variational parameters leads to 
$u_{\bm k}=\frac{1}{\sqrt{2}}\Bigl(1+\frac{\xi_{\bm k}}{\sqrt{\smash[b]{\mathstrut \xi^2\hspace{-3pt}{}_{\bm k}+\Delta^2\hspace{-3pt}{}_{\bm k}}}}\Bigr){\rule{0pt}{8pt}}^{1/2}$, 
$v_{\bm k}=\frac{1}{\sqrt{2}}\Bigl(1-\frac{\xi_{\bm k}}{\sqrt{\smash[b]{\mathstrut \xi^2\hspace{-3pt}{}_{\bm k}+\Delta^2\hspace{-3pt}{}_{\bm k}}}}\Bigr){\rule{0pt}{8pt}}^{1/2}$, where $\Delta_{\bm k}\equiv-\frac{2}{3}\sum_{{\bm k}'}U_{{\bm k},{\bm k}'}u_{{\bm k}'}v_{{\bm k}'}$ is the order parameter characterizing the Cooper triple condensation.
% The minimization of $E_0$ with respect to the variational parameters leads to 
% $u_{\bm k}=1/\sqrt{2}(1+\xi_{\bm k}/\sqrt{\smash[b]{\mathstrut \xi^2\hspace{-3pt}{}_{\bm k}+\Delta^2\hspace{-3pt}{}_{\bm k}}}){\rule{0pt}{8pt}}^{1/2}$, 
% $v_{\bm k}=1/\sqrt{2}(1-\xi_{\bm k}/\sqrt{\smash[b]{\mathstrut \xi^2\hspace{-3pt}{}_{\bm k}+\Delta^2\hspace{-3pt}{}_{\bm k}}}){\rule{0pt}{8pt}}^{1/2}$, where $\Delta_{\bm k}\equiv-\frac{2}{3}\sum_{{\bm k}'}U_{{\bm k},{\bm k}'}u_{{\bm k}'}v_{{\bm k}'}$ is the order parameter characterizing the Cooper triple condensation.
$\Delta_{\bm k}$ can also be expressed as
\begin{align}
    \Delta_{\bm k}=-\frac{2}{3}\sum_{{\bm k}'}U_{{\bm k},{\bm k}'}\bigl\langle \hat{c}^\dag\hspace{-5pt}{}_{{\bm k}',{\rm r}} \hat{c}^\dag\hspace{-5pt}{}_{\hat{R}_0{\bm k}',{\rm g}} \hat{c}^\dag\hspace{-5pt}{}_{\hat{R}_0{}^2{\bm k}',{\rm b}} \bigr\rangle,
    \label{eq:Delta}
\end{align}
which breaks the global U(1) symmetry.
$\Delta_{\bm k}$ is the expectation value of fermionic (Grassmann odd) operators, which is in contrast to the BCS ground state.
We note that the nonzero expectation value of the Grassmann-odd operators
$\bigl\langle \hat{c}^\dag\hspace{-3pt}{}_{{\bm k}',{\rm r}} \hat{c}^\dag\hspace{-3pt}{}_{\hat{R}_0{\bm k}',{\rm g}} \hat{c}^\dag\hspace{-3pt}{}_{\hat{R}_0{}^2{\bm k}',{\rm b}} \bigr\rangle\equiv\bigl\langle \Psi_T\bigl| \hat{c}^\dag\hspace{-3pt}{}_{{\bm k}',{\rm r}} \hat{c}^\dag\hspace{-3pt}{}_{\hat{R}_0{\bm k}',{\rm g}} \hat{c}^\dag\hspace{-3pt}{}_{\hat{R}_0{}^2{\bm k}',{\rm b}} \bigr|\Psi_T\bigr\rangle$
% $\langle c_{{\bm k}',{\rm r}}^\dag c_{\hat{R}_0{\bm k}',{\rm g}}^\dag c_{\hat{R}_0{}^2{\bm k}',{\rm b}}^\dag \rangle\equiv\langle \Psi_T| c_{{\bm k}',{\rm r}}^\dag c_{\hat{R}_0{\bm k}',{\rm g}}^\dag c_{\hat{R}_0{}^2{\bm k}',{\rm b}}^\dag |\Psi_T\rangle$
can be realized due to the Grassmann-odd feature of $|\Psi_T\rangle$ as well as the anti-commutation relation of $\hat{F}_{\bm{k},\hat{R}}$ with additional internal degrees of freedom given by Eq.~\eqref{eq:acomm2}.
From Eq. \eqref{eq:Energy_in_Variational_method}, we obtain
\begin{align}
    E_0
    =\frac{3}{2}\sum_{\bm k}\xi_{\bm k}\Biggl(1-\frac{\xi_{\bm k}}{\sqrt{\smash[b]{\mathstrut \xi^2\hspace{-5pt}{}_{\bm k}+\Delta^2\hspace{-5pt}{}_{\bm k}}}}\Biggr)
    +\frac{1}{4}\sum_{\bm k}\sum_{{\bm k}'}U_{{\bm k},{\bm k}'}\frac{\Delta_{\bm k}}{\sqrt{\smash[b]{\mathstrut \xi^2\hspace{-5pt}{}_{\bm k}+\Delta^2\hspace{-5pt}{}_{\bm k}}}}\frac{\Delta_{{\bm k}'}}{\sqrt{\smash[b]{\mathstrut \xi^2\hspace{-5pt}{}_{{\bm k}'}+\Delta^2\hspace{-5pt}{}_{{\bm k}'}}}}.
    \label{eq:energy(weak_coupling)}
\end{align}
%In the present case, we assume the following
To proceed 
%the calculations 
further, we consider the simplified coupling 
\begin{align}
    U_{{\bm k},{\bm k}'}=-U_0\theta(\Lambda-\abs{\xi_{\bm k}})\theta(\Lambda-\abs{\xi_{{\bm k}'}}),
    \label{eq:U}
\end{align}
where $U_0$ is the positive constant and $\Lambda$ is an energy cutoff
% $\Lambda$ 
so that the interaction acts only on the three fermions with momentum close to the FS. 
$\Lambda$ corresponds to the Debye frequency of conventional BCS superconductors with phonon-mediated interactions.
Note that $U_0$ and $\Lambda$ in the case of fermion-mediated interaction associated with an additional fermionic bath is discussed in Ref.~\cite{Akagami2021}. 
Substituting Eq. \eqref{eq:U} to Eq. \eqref{eq:Delta}, we obtain $\Delta_{\bm k}=\Delta\theta(\Lambda-\abs{\xi_{\bm k}})$, where $\Delta$ is the amplitude of the order parameter.
In the weak coupling limit ($\Delta\ll\Lambda$ and $\mu\simeq E_{\rm F}$), as in BCS theory, $\Delta$ can be obtained analytically as
\begin{align}
    \Delta\approx2\Lambda\exp\biggl(-\frac{3}{2\rho(0)U_0}\biggr),
\end{align}
where 
%$\rho(\omega)=\frac{1}{(2\pi)^2}(2m)^{3/2}(\omega+E_{\rm F})^{1/2}$
$\rho(\omega)=\sum_{\bm{k}}\delta(\omega-\xi_{\bm{k}})$
is the density of states as a function of the single-particle energy $\omega$. We used 
% \begin{align}
%     \Delta&=\frac{2}{3}U_0\sum_{\bm{k}}u_{\bm{k}}v_{\bm{k}}
%     =\frac{2}{3}U_0\int d\omega \sum_{\bm{k}}\frac{\Delta}{2\sqrt{\xi_{\bm{k}}^2+\Delta^2}}\delta(\omega-\xi_{\bm{k}})
%     =\frac{2}{3}U_0\int d\omega \sum_{\bm{k}}
%     \frac{\Delta}{2\sqrt{\omega^2+\Delta^2}}
%     \delta(\omega-\xi_{\bm{k}})
%     \cr
%     &=\frac{2}{3}U_0\int d\omega 
%     \frac{\Delta}{2\sqrt{\omega^2+\Delta^2}}
%     \sum_{\bm{k}}\delta(\omega-\xi_{\bm{k}})
%     =\frac{2}{3}U_0\int d\omega\frac{\Delta\rho(\omega)}{2\sqrt{\omega^2+\Delta^2}}\simeq\frac{2}{3}U_0\rho(0)\Delta\int_{-\Lambda}^{\Lambda}\frac{d\omega}{2\sqrt{\omega^2+\Delta^2}}\cr
%     1&\simeq\frac{2}{3}U_0\rho(0)\int_0^{\Lambda}\frac{d\omega}{\sqrt{\omega^2+\Delta^2}}\simeq \frac{2}{3}U_0\rho(0)\ln\left(\frac{2\Lambda}{\Delta}\right)
%     \rightarrow \Delta\simeq 2\Lambda\exp(-\frac{3}{2\rho(0)U_0})
% \end{align}
\vspace{-8pt}
\begin{align}
    \Delta&
    % =\frac{2}{3}U_0\sum_{\abs{\xi_{\bm k}}\leq\Lambda}u_{\bm k}v_{\bm k}
    =\frac{2}{3}U_0\sum_{\abs{\xi_{\bm k}}\leq\Lambda}\frac{\Delta}{2\sqrt{\smash[b]{\mathstrut \xi^2\hspace{-5pt}{}_{{\bm k}}+\Delta^2}}}
    \rightarrow
    1=\frac{2}{3}U_0\rho(0)\int_{-\Lambda}^\Lambda \frac{\dd{\xi_{\bm k}}}{2\sqrt{\smash[b]{\mathstrut \xi^2\hspace{-5pt}{}_{{\bm k}}+\Delta}^2}}
    \simeq\frac{2}{3}U_0\rho(0)\ln\biggl(\frac{2\Lambda}{\Delta}\biggr).
\end{align}
From Eqs. \eqref{eq:energy(weak_coupling)} and \eqref{eq:U}, the ground-state energy can be expressed as
\begin{align}
    E_0\simeq E_0^{\rm FS}+\frac{3}{2}\sum_{\abs{\xi_{\bm k}}\leq\Lambda}\abs{\xi_{\bm k}}\qty(1-\frac{\abs{\xi_{\bm k}}}{\sqrt{\smash[b]{\mathstrut \xi^2\hspace{-5pt}{}_{\bm k}+\Delta^2}}})-\frac{9}{4}\frac{\Delta^2}{U_0},
\end{align}
where we divide the summation in Eq.~\eqref{eq:energy(weak_coupling)} as $\sum_{\bm k}=\sum_{\abs{\xi_{\bm k}}>\Lambda}+\sum_{\abs{\xi_{\bm k}}\leq\Lambda}$. The summation $\sum_{\abs{\xi_{\bm k}}>\Lambda}$ 
contributes only to the noninteracting Fermi-gas energy
%gives the Fermi energy 
$E_0^{\rm FS}$ because of $U_{{\bm k},{\bm k}'}=0$ there. On the other hand, the summation $\sum_{\abs{\xi_{\bm k}}\leq\Lambda}$ is responsible for the effect of interaction near the FS.
The difference between the ground-state energies of the Cooper triple condensed state and the normal Fermi degenerate state is given by\vspace{-8pt}
\begin{align}
    E_0-E_0^{\rm FS}
    \simeq -3\rho(0)\Lambda^2\exp(-\frac{3}{\rho(0)U_0})<0
    \label{eq:condensatio_energy}
\end{align}
% Thus, whenever $U_0$ is nonzero, 
Thus, whenever $U_0$ is nonzero, 
the ground-state energy of the Cooper triple condensed state is lower than the ground-state energy of the normal Fermi gas. Eq. \eqref{eq:condensatio_energy} can also be regarded as the condensation energy of the Cooper triple, as in the BCS ground-state with the condensation of Cooper pairs~\cite{BCS}.

We briefly note that our variational method does not necessarily describe an exact ground state as is the case with the usual BCS variational method for superconductors.
% as in the usual BCS variational wave function for superconductors.
Indeed, there would be a way to improve the ansatz by enlarging the variational-parameter space.
Nevertheless, our result suggests the macroscopic occupation of Cooper triples at zero center-of-mass momentum (i.e., $F_{\bm{k},\hat{R}}^\dag F_{\bm{k}',\hat{R}}^\dag F_{\bm{k}'',\hat{R}}^\dag\cdots|0\rangle$) can be possible in the present system.
A similar discussion associated with the validity of the quartet BCS theory 
% in the comparison with
by comparison with
the quartet condensation model in finite nuclei has been made in Ref.~\cite{Baran}.

\section{Discussion: Effects of the Fermi-surface anisotropy in the lattice model}
While we have examined the possibility of Cooper-triple condensation in the low-filling case where the FS anisotropy is negligible,
such a deformation would be important for TPTC in the high-density regime.
The FS anisotropy originates from higher-order derivative terms of $\xi_{\bm{k}}$ as
\begin{align}
\label{eq:xi}
    \xi_{\bm{k}}=tk^2-\frac{t}{12}(k_x^4+k_y^4)-\mu +O(k^6),
\end{align}
for the two-dimensional square lattice.
For convenience, we define the isotropic Fermi energy as $E_{\rm F}\equiv\mu=tk_{\rm F}^2$ (note that this is correct only in the low-filling limit) as in the previous section and discuss effects of the second term in Eq.~(\ref{eq:xi}). 
%Next, we quantitatively investigated by numerical experiments 
%the extent to which a TPTC can be formed when even the energy %cutoff $\Lambda$ is taken into account.
\begin{figure}[t]
    \begin{center}
    \includegraphics[width=9cm]{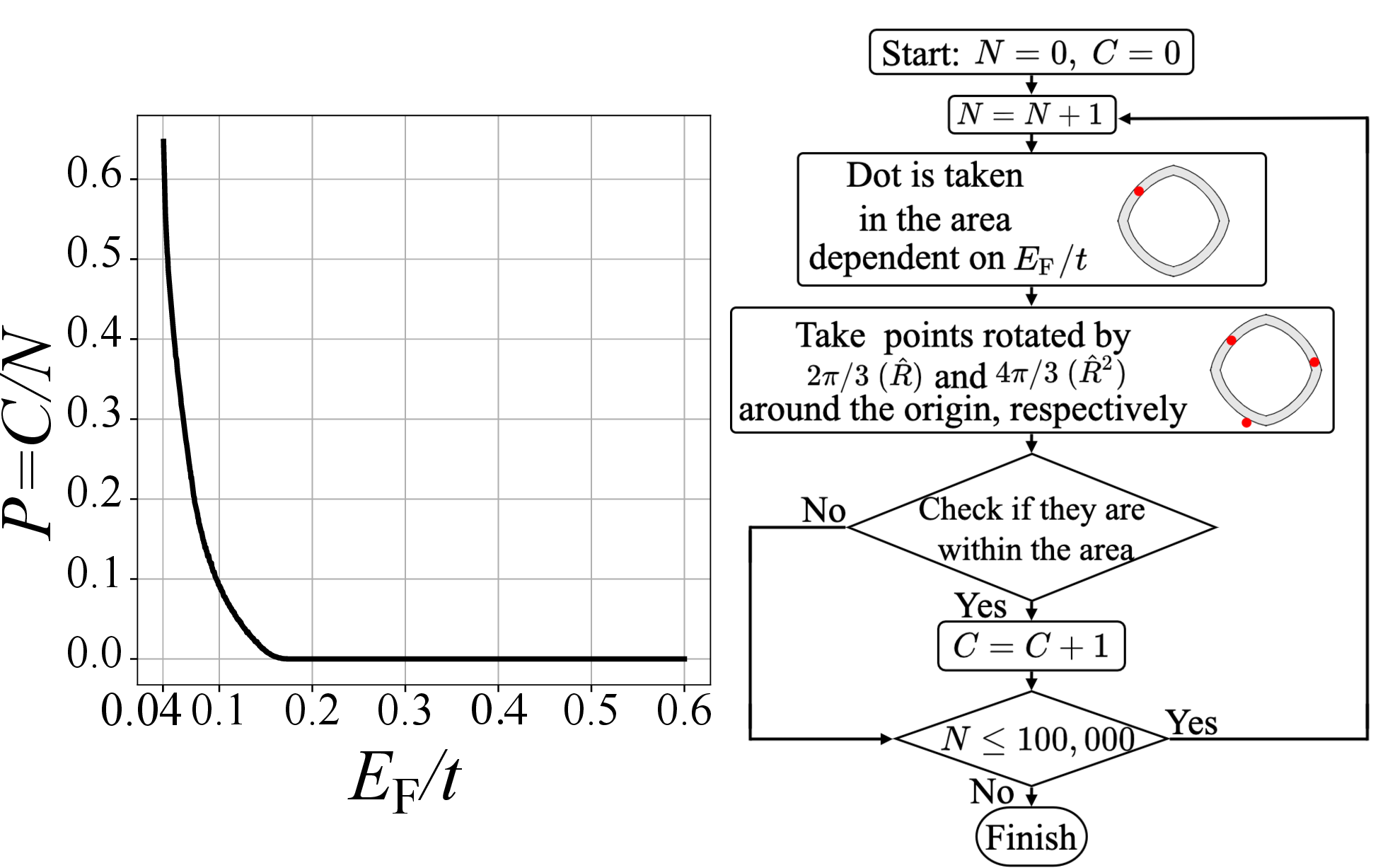}
    \caption{Probability $P$ for the realization of TPTC 
    % (i.e., the configuration of condensed Cooper triples)
    as a function of $E_{\rm F}/t$. In this plot, $\Lambda/t=0.04$ is used. 
    In the right side, the flow chart for the evaluation of $P$ is presented (see also Eq.~\eqref{eq:P}).
    }
    \label{fig:3}
    \end{center}
\end{figure}
To see how TPTC ansatz can be justified even in the presence of the FS anisotropy,
we evaluate the momentum configuration of three particles by defining
\begin{align}
\label{eq:P}
    P=\frac{C}{N}\equiv\frac{\sum_{\bm{k}}\theta(\Lambda-|\xi_{\bm{k}}|)\theta(\Lambda-|\xi_{\hat{R}\bm{k}}|)
    \theta(\Lambda-|\xi_{\hat{R}^2\bm{k}}|)}{\sum_{\bm{k}}\theta(\Lambda-|\xi_{\bm{k}}|)},
\end{align}
where $C$ (numerator) corresponds to the momentum space of TPTC within the Debye-frequency-like energy shell (i.e., $E_{\rm F}\pm\Lambda$) for the three-body interaction and
$N=\sum_{\bm{k}}\theta(\Lambda-|\xi_{\bm{k}}|)$ (denominator) 
is the configuration number where at least one fermion with $\gamma={\rm r}$ is inside of that energy shell.
In this regard, a large $P$ indicates that TPTC frequently appears in the momentum space.
We have summarized the flow chart for evaluating $P$ by using the Monte Carlo sampling with respect to the momentum integration in Fig.~\ref{fig:3}. 
The left panel of Fig.~\ref{fig:3} shows $P$ as a function of $E_{\rm F}/t$ ($\equiv \mu/t$) in Eq.~(\ref{eq:xi}), where $\Lambda/t=0.04$ is employed.
We find that TPTC are formed frequently in the region where $E_{\rm F}/t$ is sufficiently small, namely, in the low-filling regime. With increasing $E_{\rm F}/t$, $P$ decreases exponentially, 
%indicating that fluctuation effects will be more important.
but fluctuation effects will take over. One may expect the {\it formation} of Cooper triples or in-medium trimer states due to the interactions or cluster hopping even in the higher-filling case~\cite{Gotta2022,Chetcuti}.
While the reduction of $P$ in the higher-filling region does not immediately suppress the possibility of condensed Cooper triples, one should seriously examine such a fluctuation effect beyond the BCS-like state discussed in the previous section.
%On the other hand,  

\section{Conclusion}
%In conclusion, we have considered how the Cooper triple condensates in a three-component Fermi system by a variational method derived from the analogy of BCS theory.
We have discussed the condensation of Cooper triples in three-component fermionic systems by using a variational method based on the analogue of the BCS theory.
We have found that Cooper triples can macroscopically occupy the zero COM momentum state and form the condensate.
In such a nontrivial condensation, the internal degrees of freedom associated with constituent fermions in Cooper triples (i.e., relative momenta) play a crucial role in forming the condensate without conflicting with Pauli-exclusion principle.
We have also discussed effects of the FS anisotropy in the two-dimensional Hubbard model.
While TPTC (characterizing the configuration of condensed Cooper triples) within the interaction energy shell can frequently appear in the low-filling regime, 
other kinds of configurations 
%corresponding to 
accompanied by fluctuations are expected
to be dominant in the higher-filling case.

While we consider the two-dimensional system, 
it is an interesting topic to see whether the Mermin--Wagner--Hohenberg theorem~\cite{MW,Hohenberg} holds or not in such an exotic Fermi condensation.  
% Also, it is worth considering the nature of the excited Cooper triples and the formation of Cooper triples when two- and three-body interactions coexist. It would also be interesting to consider other lattices, such as honeycomb lattices~\cite{Li}.
% We are also interested in the application of our approach to dense quark matter.
% \\
Also, it is worth considering the nature of excited Cooper triples, the formation of Cooper triples when two- and three-body interactions coexist, to consider other lattices, such as honeycomb lattices~\cite{Li}, and application of our approach to dense quark matter~\cite{Ring,Baym}.\\

\vspace*{0.2cm}
This work was in part supported by Grants-in-Aid for Scientific Research provided by JSPS
through No.\ 18H05406.

\end{document}